\RequirePackage{lineno} 
\documentclass[aps,prc,superscriptaddress,nofootinbib,floatfix,twocolumn]{revtex4}

\usepackage{color}

\usepackage{graphicx} 
\usepackage{hyperref}
\usepackage{MnSymbol}
\usepackage{amsmath} 
\usepackage{gensymb} 
\usepackage{xspace}
\usepackage{ulem}

\newcommand{\PbO}         {\mbox{Pb--O}\xspace}
\newcommand{\PbNe}         {\mbox{Pb--Ne}\xspace}
\newcommand{\Six}         {$\sqrt{s_{_\mathrm{NN}}}~=~68.5$~Ge\kern-.1emV\xspace}

\newcommand{\Abinitio}{\textit{ab initio}\xspace}

\newcommand{\Oxygen}{$^{16}$O\xspace}
\newcommand{\Neon}{$^{20}$Ne\xspace}

\usepackage{color}
\definecolor{darkgreen}{RGB}{20,200,20}
\definecolor{darkgray}{RGB}{128,128,128}

\begin{document} 

\title{Signature of the $\alpha$-clustering structure of Light Nuclei in Relativistic Nuclear Collisions}
\author{Zhiyong Lu} 
\affiliation{China Nuclear Data Center and China Institute of Atomic Energy, Beijing 102413, China}
\affiliation{Niels Bohr Institute, Jagtvej 155A, 2200 Copenhagen, Denmark}
\author{Mingrui Zhao} 
\affiliation{China Nuclear Data Center and China Institute of Atomic Energy, Beijing 102413, China}
\author{Emil Gorm Dahlb\ae k Nielsen}
\affiliation{Niels Bohr Institute, Jagtvej 155A, 2200 Copenhagen, Denmark}
\author{Xiaomei Li} 
\email[co-corresponding author: ]{lixm@cern.ch}
\affiliation{China Nuclear Data Center and China Institute of Atomic Energy, Beijing 102413, China}
\author{You Zhou}
\email[corresponding author: ]{you.zhou@nbi.ku.dk}
\affiliation{Niels Bohr Institute, Jagtvej 155A, 2200 Copenhagen, Denmark}

\date{\today}

\begin{abstract}
The “imaging-by-smashing” technique has been developed recently in relativistic nuclear collisions. By smashing heavy nuclei at RHIC and the LHC and analyzing the anisotropic expansion (flow) of the final state produced particles, unique information on the structure of the collided nuclei has been obtained. Existing efforts primarily focus on the colliding mode of heavy nuclei collisions. In contrast, nuclear structure studies with collisions of light nuclei and the fixed target mode, despite their significant impact and broad interest, have not been thoroughly explored. In this Letter, we investigate the $\alpha$-clustering signature of $^{20}$Ne and $^{16}$O in the fixed-target $^{208}$Pb--$^{20}$Ne and $^{208}$Pb--$^{16}$O collisions at $\sqrt{s_{_\mathrm{NN}}}$ = 68.5 GeV, using the parton transport model AMPT. The results of two- and four-particle cumulants of anisotropic flow demonstrate a robust $\alpha$-clustering signature that persists regardless of the complex dynamic evolution of the created systems. This study highlights the significant impact of the LHCb SMOG (SMOG2) project in discovering the $\alpha$-clustering signature of light nuclei at relativistic energies.
\end{abstract}
\maketitle


\section{Introduction}
\label{sec:intro}
The primary goal of ultra-relativistic nuclear collisions is to recreate the quark-gluon plasma (QGP), a state of matter believed to have existed in the early universe, and to investigate its properties~\cite{STAR:2005gfr,PHENIX:2004vcz,ALICE:2022wpn}.
Precise determinations of the QGP properties and their time evolution rely on an accurate description of the initial conditions of relativistic nuclear collisions. 
The initial conditions are determined by the overlap of colliding nuclei, which are often described by the Woods-Saxon profile~\cite{dEnterria:2020dwq}. 
Recent studies~\cite{Giacalone:2021udy,Bally:2021qys,Jia:2022qgl,Zhang:2021kxj,Zhao:2022uhl} suggest that the nuclear geometry could significantly affect the shape of the initial conditions, impacting many final experimental observations.
Among these experimental observables, one of the most important observables is the anisotropic flow~\cite{Ollitrault:1992bk}. This phenomenon is caused by the initial spatial anisotropy being transferred to the momentum anisotropy in the final state via the pressure gradients within the created medium.
The anisotropic flow can be characterised by the Fourier coefficients $v_n$ of the azimuthal particle distribution
\begin{equation}
    \frac{dN}{d\varphi} \propto 1+2\sum\limits_{n=1}\limits^{\infty}{v_n\cos[n(\varphi-\Psi_n)]},
    \label{eq:FourierSeries}
\end{equation}
where $\varphi$ is the azimuthal angle of the final state produced particle, $\Psi_n$ is the $n^{\rm th}$-order symmetry plane, and $v_n$ is the flow coefficient.
The size of flow coefficients has been used as a powerful probe of the QGP's properties~\cite{ALICE:2011ab,ATLAS:2012at,CMS:2013wjq,ALICE:2016ccg}, i.e., shear and bulk viscosities, and also allows direct access to the initial conditions~\cite{Gardim:2011xv,Niemi:2012aj,Gardim:2014tya,Li:2021nas}.
By smashing together heavy nuclei at RHIC and the LHC and analyzing the anisotropic flow of the final state emitted particles, unique information on the initial conditions shortly after the collisions and the structure of the collided nuclei before the collision could be obtained~\cite{Jia:2021qyu}.
This “imaging-by-smashing” technique has been extensively applied in various experimental programs, such as the isobar runs of Zirconium-Zirconium ($^{96}$Zr--$^{96}$Zr), Ruthenium-Ruthenium ($^{96}$Ru--$^{96}$Ru) collisions~\cite{Zhang:2021kxj}, and Uranium-Uranium ($^{238}$U--$^{238}$U) collisions at RHIC~\cite{STAR:2015mki,STAR:2024wgy}, as well as the Xenon-Xenon ($^{129}$Xe--$^{129}$Xe) collisions~\cite{ALICE:2021gxt,ATLAS:2022dov,ALICE:2024nqd} at the LHC. 
These studies have provided important insights into nuclear structures, including the evidence of quadrupole deformation~\cite{STAR:2015mki,STAR:2024wgy,ALICE:2021gxt,ATLAS:2022dov,ALICE:2024nqd} and the indication of triaxiality~\cite{Bally:2021qys,Ryssens:2023fkv,Zhao:2024lpc}. They have also demonstrated the significant role that these structures play in shaping the initial conditions and subsequent QGP evolution~\cite{ALICE:2021gxt,STAR:2024wgy}.

The “imaging-by-smashing” technique has primarily been used on heavy nuclei, but its extension to light nuclei also brings tremendous opportunities. Among the light nuclei, \Oxygen and \Neon are of great interest to study, as they are predicted to have novel shapes due to the $\alpha$-clustering phenomena~\cite{gamow1930mass}, suggested by low-energy nuclear physics, i.e., shown by the \Abinitio calculations~\cite{Giacalone:2024luz}.
Previous studies suggest that the $\alpha$-clustering structure could be investigated in the \Oxygen--\Oxygen and \Neon--\Neon collisions at RHIC~\cite{Huang:2023viw} and the LHC~\cite{ALICE-PUBLIC-2021-004}.
So far, the primary focus has been on the colliding mode of the ``smashing experiments", the potential of using fixed-target experiments to image the nuclear structure has not yet been explored thoroughly. 

The LHCb System for Measuring Overlap with Gas (SMOG) project~\cite{2707819}, where the LHC beams collide with gas targets at rest, has collected data on $^{208}$Pb--$^{20}$Ne (\PbNe in short) in the LHC Run 2 and has been collecting more \PbNe samples during the ongoing LHC Run 3. 
Additional $^{208}$Pb--$^{16}$O (\PbO in short) collisions were also proposed to be operated at Run 3, with a centre of mass energy of around 70 GeV.
The data taken in SMOG and its updated version SMOG2 provides unique opportunities to discover $\alpha$-cluster in \Neon and \Oxygen.
Inspired by the “imaging-by-smashing" technique in ultra-relativistic nuclear collisions, the anisotropic flow in these systems might be a powerful tool for discovering $\alpha$-cluster in light nuclei.
Notably, the centre of mass energy at the SMOG is significantly lower than the LHC and the top RHIC energies, where the “imaging-by-smashing" technique is valid. 
Whether the approach is still valid at this low energy with \PbNe and \PbO collisions is to be investigated.
Therefore, theoretical works supporting the imaging of the $\alpha$-cluster of \Neon and \Oxygen at SMOG (SMOG2) are still necessary.

A recent study employing a hydrodynamic model~\cite{Giacalone:2024ixe}, which implemented the initial conditions with inputs from \Abinitio, was conducted to investigate the $\alpha$-clustering structure in \Neon and \Oxygen. However, the prerequisites for applying hydrodynamics are not entirely clear. Competing models, such as parton transport models, have not been investigated at the moment. In small and medium systems, transport models feature limited parton interactions, whereas hydrodynamic models use sufficient interactions in the created medium. These fundamental differences highlight the necessity of conducting a relevant study using transport models. Additionally, non-negligible short-range few-particle azimuthal correlations, known as non-flow effects, could significantly affect such studies. These non-flow effects were not well simulated with hydrodynamic models but could be studied using transport models such as AMPT.

In this Letter, the nuclear structures of \Neon and \Oxygen have been investigated via the study of anisotropic flow, utilizing the transport model AMPT with various considerations for the nuclear structures of \Neon and \Oxygen. The comparisons between these results allow for probing the $\alpha$-clustering signature. The paper is organised as follows: first, the model setups in this study are introduced in Sec.\ref{sec:model}. Section \ref{sec:obser} defines the observables used. The results are discussed in Sec.\ref{sec:result}, and conclusions are provided in Sec.\ref{sec:conclusion}.

\section{model setups}
\label{sec:model}

The simulations of ultra-relativistic \PbNe and \PbO collisions at \Six are implemented using A Multi-Phase Transport (AMPT) model~\cite{Lin:2004en} with a string melting scenario. The AMPT model consists of several processes. First, nucleon spatial and momentum distributions of the initial nucleons within the nuclei are given by the HIJING model~\cite{Wang:2000bf}.
These nucleon distributions are then transferred to parton distribution. The interactions among partons are controlled by Zhang's Parton Cascade (ZPC) model~\cite{Zhang:1997ej}.
In this ZPC model, parton-scattering cross-section $\sigma$ depicts the dynamic expansion of the QGP phase
\begin{equation}
    \sigma = \frac{9\pi\alpha_s^2}{2\mu^2},
    \label{eq:parton-scattering cross-sections}
\end{equation}
where $\alpha_s$ is the QCD coupling constant, $\mu$ is the screening mass that decides the strength of parton interactions. After ZPC, the hadronisation process takes place by employing a coalescence model~\cite{Chen:2005mr}. Finally, the hardon interactions are described by the ART model~\cite{Li:1995pra}.

In this analysis, the nucleon distributions in HIJING for \Oxygen and \Neon are configured according to the state-of-the-art \Abinitio from Nuclear Lattice Effective Field Theory (NLEFT)~\cite{Giacalone:2024luz}, with only positive-sign nuclear configurations considered. Note that the sign problem of NLEFT has been thoroughly discussed in Ref.~\cite{Giacalone:2024luz}, which showed that it is acceptable to assign only positive weight. The NLEFT calculations show the ground state of \Neon and \Oxygen nuclei with the $\alpha$-clustering structures~\cite{Giacalone:2024luz}. For the comparisons, the baselines were selected where the nucleon distributions of \Oxygen and \Neon follow Woods-Saxon profiles in a three-parameter Fermi (3pF) model~\cite{DeVries:1987atn}
\begin{equation}
    \rho(r) = \frac{\rho_0(1+w r^2/R_0^2)}{1+e^{(r-R_0)/a}}.
    \label{WoodsSaxon_3pF}
\end{equation}
Here, $R_0$ denotes the nuclear radius, $a$ is the depth of the nuclear surface, and $w$ is the weight parameter. The nuclear saturation density $\rho_0$ matches the integral of the distribution to the number of nucleons. The values of these parameters, listed in Tab.~\ref{tab:3pFmodel}, are chosen based on the root-mean-square nuclear charge radius~\cite{Angeli:2013epw}.

\begin{table}[hbt]
    \caption{3pF model configurations}
    \label{tab:3pFmodel}
    \begin{center}
    \begin{tabular}{ccccc}
    \hline
    nucleus & $\rho_0$ (fm$^{-3}$) & $w$ (fm) & $R_0$ (fm) & $a$ (fm)\\
    \hline
	\Oxygen & 0.0103 & -0.051 & 2.608 & 0.513 \\
	\Neon & 0.0090 & -0.168 & 2.791 & 0.698 \\
    \hline
    \end{tabular}
    \end{center}
\end{table}

For $^{208}$Pb, the nucleus is configured in a two-parameter Fermi (2pF) model~\cite{DeVries:1987atn}
\begin{equation}
    \rho(r) = \frac{\rho_0}{1+e^{(r-R_0)/a}},
    \label{WoodsSaxon_2pF}
\end{equation}
where $R_0=6.624$ fm, $a=0.549$ fm~\cite{DeVries:1987atn}.

How the $\alpha$-clustering signature is affected by the dynamic evolution of the systems can be explored by changing the strength of parton interactions in the ZPC model. Instead of the default value of the screening mass $\mu = 2.265~\mathrm{fm^{-1}}$ that is commonly used~\cite{Lin:2004en}, an additional simulation was conducted with $\mu = 226.5~\mathrm{fm^{-1}}$, which effectively turns off the partonic interaction in ZPC. The comparison of the results from the two simulations will be used to investigate the impact of nuclear structure through different dynamic evolutions of the systems.

\section{Observables}
\label{sec:obser}
Flow coefficients $v_n$ quantify the anisotropy of the final state and show great sensitivities to initial geometries~\cite{Jia:2021tzt,Zhang:2021kxj,Giacalone:2021udy,Jia:2022qgl,Magdy:2022cvt,Jia:2022qrq,Xu:2021uar,Jia:2021qyu, Lu:2023fqd}.
Usually $v_n$ can be calculated with two-particle correlation~\cite{Borghini:2000sa,Borghini:2001vi,Bilandzic:2010jr,Bilandzic:2013kga,Moravcova:2020wnf}
\begin{equation}
    v_n\{2\}\equiv \sqrt{c_n\{2\}}=\langle\langle\cos n(\varphi_1-\varphi_2)\rangle\rangle^{1/2},
    \label{eq:TwoParticleCumulants}
\end{equation}
where $\varphi_1$ and $\varphi_2$ are the azimuthal angles of randomly selected two (different) particles in a single event. 
Double brackets are the average over all events, after the average over all particles in a single event.
The particles are selected from the kinematic regions of $0.2<p_{\mathrm{T}}<5.0~{\rm GeV}$ and $2.0<\eta<5.0$, which are within the LHCb acceptance.
To suppress non-flow effects, which are the contamination coming from resonances decay, jet et al., the $\eta$ region for $v_n\{2\}$ calculations was split into two sub-regions: $2<\eta<3$ and $4<\eta<5$.
The $\varphi_1$ and $\varphi_2$in eq.(\ref{eq:TwoParticleCumulants}) must be selected from these two different $\eta$ regions, respectively, and the two-particle correlations can be obtained following Eq. (9) in Ref.~\cite{Zhou:2015iba}.

In addition to the two-particle correlation, $v_n$ can also be obtained from the four-particle cumulant~\cite{Borghini:2000sa,Borghini:2001vi,Bilandzic:2010jr,Bilandzic:2013kga,Moravcova:2020wnf}
\begin{equation}
    v_n\{4\}\equiv\sqrt[4]{-c_n\{4\}}=\sqrt[4]{2\langle v_n^2\rangle^2 - \langle v_n^4\rangle},
\end{equation}
where
\begin{equation}
\begin{split}
    \langle v_n^2\rangle&=\langle\langle\cos n(\varphi_1-\varphi_2)\rangle\rangle,\\
    \langle v_n^4\rangle&=\langle\langle\cos(n\varphi_1+n\varphi_3-n\varphi_2-n\varphi_4)\rangle\rangle.
    \label{eq:FourParticleV2}
\end{split}
\end{equation}
The anisotropic flow of four-particle cumulant $v_n\{4\}$ is less sensitive to non-flow by construction~\cite{Borghini:2000sa,Borghini:2001vi}; thus, no additional non-flow suppression is applied in this study. The $v_n\{2\}$ and $v_n\{4\}$ carry opposite contributions from the flow fluctuations~\cite{Voloshin:2007pc}. Simultaneous studies on $v_n\{2\}$ and $v_n\{4\}$ bring additional information on $v_n$ fluctuations, which is essential in constraining the initial conditions and potentially bringing independent information on the structure of the colliding nuclei. The above-mentioned multi-particle correlation observables can be obtained using the {\it Generic~Algorithm}~\cite{Moravcova:2020wnf}.

\section{Results}
\label{sec:result}
\begin{figure}[!htb]
    \begin{center}
      \includegraphics[width=\linewidth]{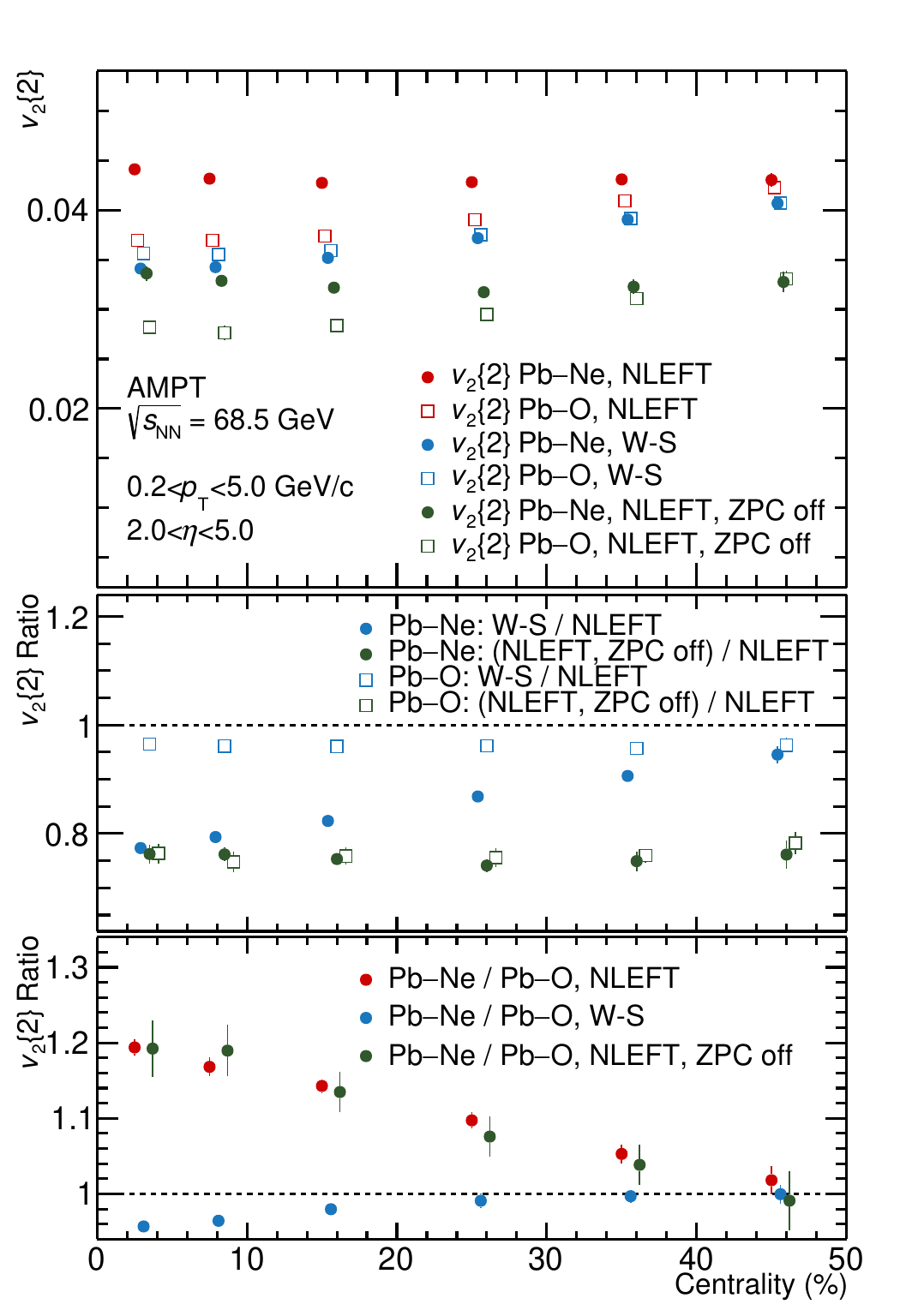}
      \caption[.]{Centrality dependence of $v_2\{2\}$ in \PbNe and \PbO collisions at $\sqrt{s_{_\mathrm{NN}}}$ = 68.5 GeV in AMPT. The X-axis of different markers is shifted for visibility. Red markers: \Neon and \Oxygen nucleus are described by NLEFT calculations ($\alpha$-cluster). Blue markers: \Neon and \Oxygen nuclei are described by the Woods-Saxon profile. Green markers: \Neon and \Oxygen nucleus are described by NLEFT calculations, and the ZPC in AMPT is turned off. }
      \label{fig:v22}
    \end{center}
\end{figure}

The centrality dependence of $v_2\{2\}$ in \PbNe and \PbO collisions at $\sqrt{s_{_\mathrm{NN}}}$ = 68.5 GeV are presented in the top panel of Fig.~\ref{fig:v22}.
The ratios of different AMPT configurations and collision systems are presented in the middle and bottom panels, respectively.
In \PbNe collisions, $v_2\{2\}$ with NLEFT configuration (solid red circles) is significantly larger than the one with Woods-Saxon configuration (solid blue circles) in 0--50\% centrality range. 
The ratio of $v_2\{2\}$ in \PbNe collisions between NLEFT and Woods-Saxon configurations is less than 0.8 in the most central collisions. This could be explained by the bowling pin shape of \Neon, a shape consisting of five $\alpha$ clusters predicted by the NLEFT calculations~\cite{Giacalone:2024luz}. In most central collisions, the shape of the overlap region is largely determined by the shape of colliding nuclei. The bowling pin shape of \Neon enhances the initial eccentricity $\varepsilon_2$ and hence the $v_2\{2\}$. When the centrality is above 50\%, $v_2\{2\}$ in \PbNe collisions with NLEFT configuration agrees with the one with Woods-Saxon configuration.
This shape of the initial conditions is predominantly determined by the overlapping region, which might not be very sensitive to the shape of the colliding nuclei.
In \PbO collisions, $v_2\{2\}$ with NLEFT configuration (open red square) is slightly larger than the one with Woods-Saxon configuration (open blue square) in the 0--70\% centrality range, with the ratio between these two configurations staying around 0.95.
The tetrahedron shape~\cite{Giacalone:2024luz} of \Oxygen, a shape consisting of four $\alpha$-clusters predicted by NLEFT calculations, only results in a small increase of $v_2\{2\}$, compared to the spherical shape described by Woods-Saxon profile.
As a result, the $\alpha$-clustering signature in \PbO collisions is inconspicuous and challenging to detect.

Although significant differences of $v_2\{2\}$ between NLEFT and Woods-Saxon configurations have been observed in \PbNe collisions, which suggest the $\alpha$-clustering structure, a robust signature can be achieved by comparing the $v_2\{2\}$ in \PbNe and \PbO collisions, or by looking at the ratio of $v_2\{2\}$ from the two collision systems. Considering that the \PbNe and \PbO collisions have similar system sizes, the ratios of flow observables in \PbNe and \PbO collisions are expected to be independent of the final state effects and thus bring direct access into the nuclear structures of \Neon and \Oxygen. The ratio of $v_2\{2\}$ between the two collision systems is presented in the bottom panel of Fig.~\ref{fig:v22}.
With NLEFT configuration for \Neon and \Oxygen nuclei, the ratio of $v_2\{2\}$(\PbNe)/$v_2\{2\}$(\PbO) (red solid circle) starts from about 1.2 in the most central collisions and decreases to unity at around 50\% centrality. 
While with Woods-Saxon configuration of \Neon and \Oxygen nuclei, $v_2\{2\}$(\PbNe)/$v_2\{2\}$(\PbO) 
(blue solid circle) starts at about 0.96 at the most central collisions and stays around unity for centrality larger than 30\%.
If \Neon and \Oxygen have near $\alpha$-clustering structures on the ground state, the ratio of $v_2\{2\}$(\PbNe)/$v_2\{2\}$(\PbO) will be distinctly larger than unity in central collisions. Otherwise, the ratio will be around unity or even below unity. This is a strong signature of $\alpha$-clustering structure of \Neon and \Oxygen nuclei. It is noteworthy that this $\alpha$-clustering signature has also been confirmed by hydrodynamic calculations~\cite{Giacalone:2024ixe}, despite the dramatic difference in the dynamic evolution of the created systems compared to the AMPT study presented in this paper.

For the dynamic evolution of the created systems in the light nuclei collisions, it's still not entirely clear whether it undergoes dilute or dense parton interactions. The effects of the different dynamic evolutions could be explored by enormously enlarging the screen mass in ZPC to turn off the parton interactions. After that, smaller $v_2\{2\}$ results are shown for both \PbNe and \PbO collisions (from red to green markers in the upper panel), while their centrality dependences barely change. The decrease of $v_2\{2\}$ can be quantified in the middle panel, where the ratios of $v_2\{2\}$(NLEFT, ZPC off)/$v_2\{2\}$(NLEFT) in both \PbNe and \PbO collisions start at about 0.75 in the most central collisions and slowly increase to about 0.85 at 70\% centrality. 
The magnitudes of $v_2\{2\}$ show the strength of parton interactions in \PbNe and \PbO collisions. Future data-model comparisons could help improve understanding of the properties of the created matter in ultra-relativistic fixed target experiments at the LHC.
Most importantly, the ratio of $v_2\{2\}$(\PbNe)/$v_2\{2\}$(\PbO) with (NLEFT, ZPC off) configuration agrees with the results using the default value of $\mu$. This agreement shows that the usage of ratio observables from two colliding systems can cancel the final state effect and reflect the information from the structures of \Neon and \Oxygen directly. It confirms that $\alpha$-clustering signature in $v_2\{2\}$(\PbNe)/$v_2\{2\}$(\PbO) persists, and it is independent of the properties of the created matter at the LHC fixed target experiments. 

\begin{figure}[!htb]
    \begin{center}
      \includegraphics[width=\linewidth]{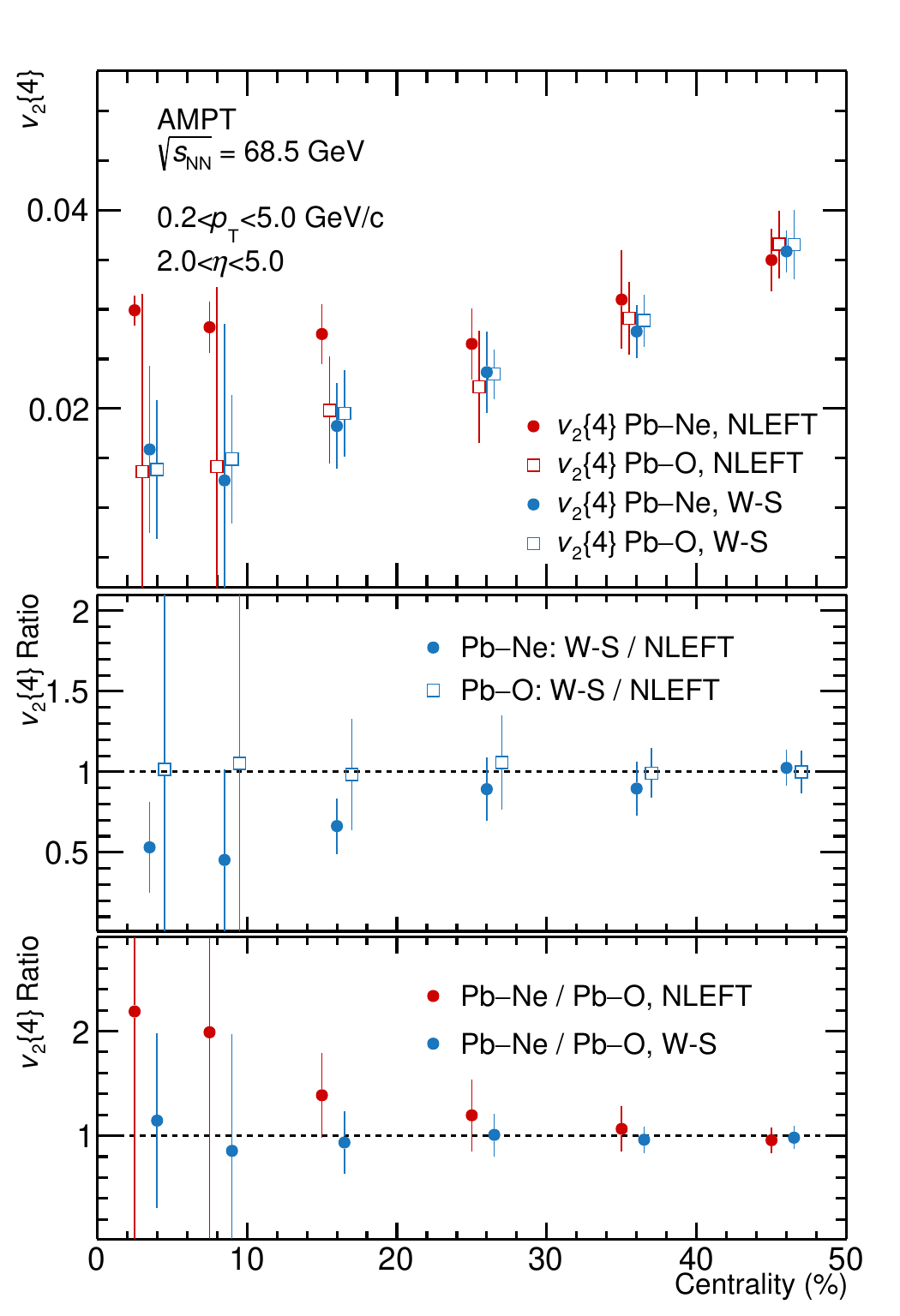}
      \caption[.]{Centrality dependence of $v_2\{4\}$ in \PbNe and \PbO collisions at $\sqrt{s_{_\mathrm{NN}}}$ = 68.5 GeV in AMPT. The X-axis of different markers is shifted for visibility. Red markers: \Neon and \Oxygen nucleus are described by NLEFT calculations ($\alpha$-cluster). Blue markers: \Neon and \Oxygen nuclei are described by the Woods-Saxon profile.}
      \label{fig:v24}
    \end{center}
\end{figure}

The non-flow effects are non-negligible in colliding systems of small and medium sizes. The $|\Delta\eta| >1$ method for $v_2\{2\}$ is expected to largely suppress the non-flow effect.
Larger $\eta$ gaps have also been examined in this AMPT study, and the possible remaining non-flow in $v_2\{2, |\Delta\eta| >1\}$ has negligible impact on the conclusions drawn from Fig.~\ref{fig:v22}.
In other words, the $\alpha$-clustering signature holds when applying larger $\eta$ gaps. Besides, flow fluctuations are also considerable in small and medium systems. It's of critical importance to study $v_2\{2\}$ and $v_2\{4\}$ simultaneously as they have different sensitivities to the flow fluctuations~\cite{Voloshin:2007pc}.
The results of $v_2\{4\}$ are presented in Fig.~\ref{fig:v24}. In \PbNe collisions, $v_2\{4\}$ with NLEFT configuration (red solid circle) is larger than the one with Woods-Saxon configuration (blue solid circle) in 0--20\% centrality range because the bowling pin shape of \Neon enhances the initial eccentricity.
This also shows the influence of $\alpha$-clustering structure of \Neon in the $v_2\{4\}$ results.
The uncertainties of $v_2\{4\}$ in \PbNe collisions with Woods-Saxon configuration are relatively large due to the nearly zero values, and more simulations will be helpful for the study of the four-particle cumulant.
In \PbO collisions, $v_2\{4\}$ with Woods-Saxon and NLEFT configurations are consistent within sizable uncertainties in central collisions.

The study using the hydrodynamic model also revealed the signature of the $\alpha$-cluster, showing a significant difference in $v_2$ in \PbNe collisions between the NLEFT and Woods-Saxon configurations~\cite{Giacalone:2024ixe}, although $c_2{4}$ was reported instead of $v_2{4}$. Despite different sensitivities to flow fluctuations, the signature of the $\alpha$-cluster is still observed when the systems undergo either hydrodynamic or only a few times partonic interactions. Note, $c_2\{4\}$ was presented in the hydrodynamic studies~\cite{Giacalone:2024ixe}s because the $c_2\{4\}$ results in the central collisions are explicitly positive, preventing the calculation of real-valued $v_2\{4\}$. In the collision system with small or medium size, the positive sign of $c_2\{4\}$, if not due to the significant non-flow effect, suggests the significant nonlinear hydrodynamic response of the final state anisotropic flow to the initial eccentricity~\cite{Zhao:2017rgg}. On the other hand, the negative sign of $c_2\{4\}$ agrees with the expectation of dilute partonic interactions in the system. The sign of $c_2\{4\}$ in \PbNe and \PbO collisions should be examined in future measurements, which eventually answers whether the created systems are dilute or dense, improving our understanding of the origin of flow in the small systems.

\section{conclusion}
\label{sec:conclusion}
By utilizing the "imaging-by-smashing" technique in fixed-target \PbNe and \PbO collisions at the LHC, the investigations on the $\alpha$-clustering signature in \Neon and \Oxygen have been performed based on the AMPT simulations. The results show that for the elliptic flow of the two-particle cumulant $v_2\{2\}$, the values are significantly larger in \PbNe collisions than in \PbO collisions when the nuclear structure configuration from NLEFT calculations is used for \Neon and \Oxygen. In contrast, the values are smaller in \PbNe collisions than those in \PbO collisions when using the nuclear structure from Woods-Saxon distributions. Similar results are observed in the study of the elliptic flow of four-particle cumulants $v_2\{4\}$, where a significantly larger $v_2\{4\}$ is observed in \PbNe collisions when the NLEFT calculations are applied. The larger $v_{2}$ observed in central Pb-Ne collisions could serve as the signature of the $\alpha$-clustering structure of \Neon and \Oxygen, which is robust against the complex dynamic evolution of the created system.

Meanwhile, the negative sign of $c_2\{4\}$, correspondingly the real-valued $v_2\{4\}$, was observed in the \PbO collisions in the presented AMPT study, while a positive $c_2\{4\}$ was previously reported in hydrodynamic model calculations. Such a distinction of $c_2\{4\}$ in \PbO collisions will provide a powerful tool for revealing the origin of flow observed in small systems, whether the systems undergo hydrodynamic evolution or only a few partonic interactions.

It should be emphasised that the LHCb SMOG2 program's ability to switch gas targets provides remarkable flexibility, enabling the exploration of a broader range of nuclear structures by smashing the ion beam at relativistic energies into the gas targets. This unique capability at LHCb SMOG2 is especially valuable given the closure of RHIC and the current challenges associated with the limited species available in the LHC ion beams.

\section{Acknowledge}
\label{sec:acknowledge}
We thank Giulia Manca and Xinli Zhao for their valuable feedback. We would also like to thank the organisers and participants of the workshop 'Light Ion Collisions at the LHC (11-15 November 2024, CERN)', where the works were discussed in the early stages.
Z. Lu, M. Zhao and X. Li are supported by the National Key Research and Development Program of China (2024YFA1610804, 2022YFA1602103, 2018YFE0104800) and Continuous-Support Basic Scientific Research Project (BJ010261223282).  Z. Lu, M. Zhao, E. G. D. Nielsen and Y. Zhou are supported by the European Union (ERC, InitialConditions), the VILLUM FONDEN (grant number 00025462), and the Independent Research Fund Denmark (DFF-Sapere Aude grant, 2023).

\bibliographystyle{utphys}
\bibliography{Reference}

\appendix

\section*{Additional results}
\label{sec:appendix}
\begin{figure}[!htb]
    \begin{center}
      \includegraphics[width=\linewidth]{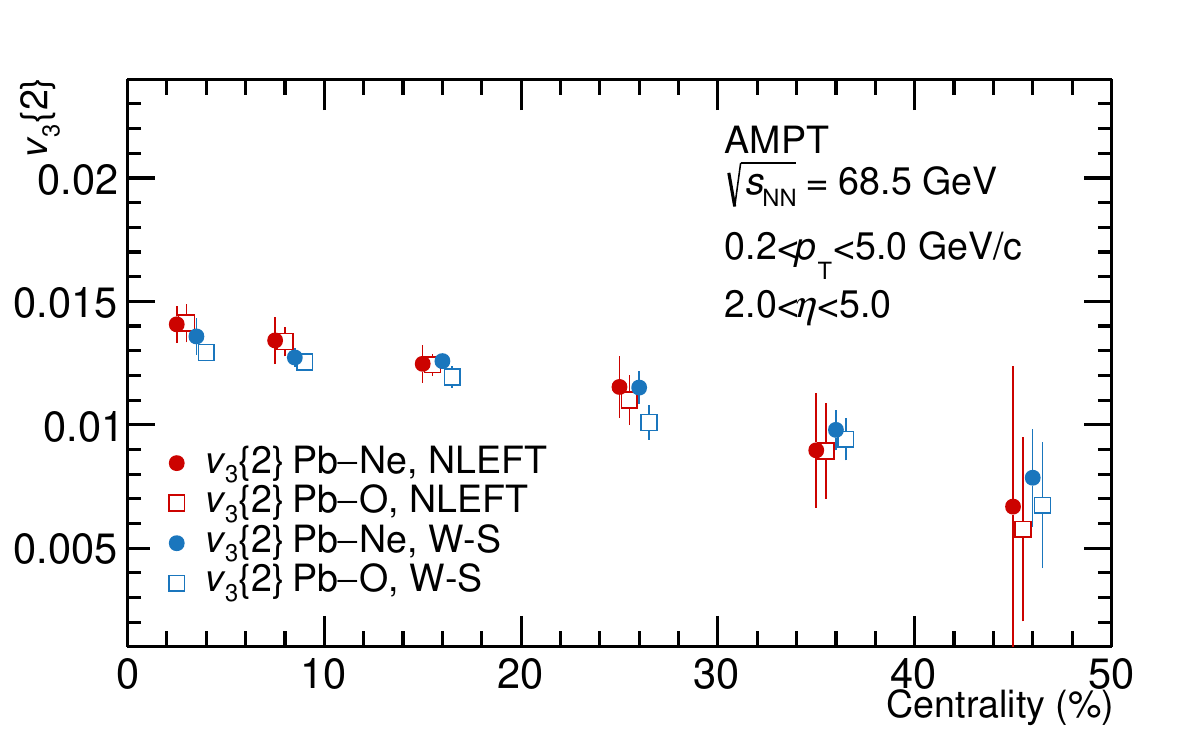}
      \caption[.]{Centrality dependence of $v_3\{2\}$ in \PbNe and \PbO collisions at $\sqrt{s_{_\mathrm{NN}}}$ = 68.5 GeV in AMPT. The X-axis of different markers is shifted for visibility. Red markers: \Neon and \Oxygen nucleus are described by NLEFT ($\alpha$-cluster). Blue markers: \Neon and \Oxygen nucleus are described by the Woods-Saxon profile.}
      \label{fig:v32}
    \end{center}
\end{figure}

The appendix presents the centrality dependence of $v_3\{2\}$ in \PbNe and \PbO collisions. Figure \ref{fig:v32} shows that $v_3\{2\}$ decreases with increasing centrality, regardless of the configurations of the nuclear structure. The differences of $v_3\{2\}$ among the four configurations are compatible within sizeable uncertainties, showing that $v_3$ is insensitive to the $\alpha$-cluster structure of \Neon and \Oxygen.

\end{document}